# Thermal-induced unfolding-refolding of a nucleocapsid COVN protein

Warin Rangubpit[1, 2], Pornthep Sompornpisut[2], R.B. Pandey[1*]

[1]School of Mathematics and Natural Sciences, University of Southern Mississippi, Hattiesburg, MS 39406, USA

[2]Center of Excellence in Computational Chemistry, Department of Chemistry, Chulalongkorn University, Bangkok 10330, Thailand

*Corresponding author, email: ras.pandey@usm.edu

Abstract: Unfolding of a coarse-grained COVN protein from its native configuration shows a linear response with increasing temperature followed by non-monotonic double peaks in its radius of gyration. The protein conforms to a random coil of folded segments in native state with increasingly tenuous and globular structures in specific temperature regimes where the effective dimensions of corresponding structures $D \approx 1.6 – 2.4$. Thermal agitation alone is not sufficient to fully eradicate its segmental folding as few local folds are found to persist around such residues as $^{65}W$, $^{110}Y$, $^{224}L$, $^{374}P$ even at high temperatures.

CoVID-19 pandemic is attracting unprecedented attention [1-5] in investigating the corona virus and its constituents. Corona virus involves a number of proteins, RNA and a huge list of crowded inter- and intra-cellular constituents in its assembly and replication. In an initial investigation even with a coarse-grained computer simulation model it is not feasible to consider all constituents that are involved in its assembly and replication. We examine the structural dynamics of a nucleocapsid (COVN) protein [6] consisting of 422 residues which plays a critical role in packaging the viral genome RNA into ribonucleocapsid and virion assembly [7-9] . For the sake of simplicity and to develop a clear understanding of the basic nature of the conformational evolution, it would be interesting to examine the structural response of a free COVN as a function of temperature before systematically including different types of proteins, solute, solvent etc. of the underlying host space.

'Protein folding' [10,11] remains an open problem despite enormous efforts for over half a century. Because of the enormity of challenges (e.g. time scale for huge degrees of freedom with all-atom approaches), coarse-graining [12-17] remains a viable choice to gain insight into the fundamental mechanism of conformational dynamics. Using a simplified yet efficient and effective coarse-grained model [18,19] , a large-scale Monte Carlo simulation is performed to study the thermal response of COVN. Our coarse-grained model has already been used to investigate structural dynamics of such proteins as histones critical in assembly of chromatin [20], lysozyme [21] and alpha-synuclein [22] key in amyloid, protein (VP40) in ebola virus [23], membrane proteins [18,19] for selective transports, etc. COVN is represented by a chain of 422 residues in a specific sequence in a cubic lattice [18,19]. Each residue interacts with surrounding residues within a range ($r_c$) with a generalized Lennard-Jones potential,

$$U_{ij} = \left[ |\varepsilon_{ij}| \left( \frac{\sigma}{r_{ij}} \right)^{12} + \varepsilon_{ij} \left( \frac{\sigma}{r_{ij}} \right)^{6} \right], r_{ij} < r_c \qquad (1)$$

where $r_{ij}$ is the distance between the residues at site $i$ and $j$; $r_c = \sqrt{8}$ and $\sigma = 1$ in units of lattice constant. A knowledge-based [12-17] residue-residue contact matrix (based on a large ensemble of protein structures in PDB) is used as input for the potential strength $\varepsilon_{ij}$ [14] in phenomenological interaction (1). With the implementation of excluded volume and limits on the covalent bond length constraints, each residue performs its stochastic movement with the Metropolis algorithm, i.e. with the Boltzmann probability *exp(-ΔE/T)* where *ΔE* is the change in energy between new and old position. Attempts to move each residue once defines unit Monte Carlo time step. All quantities are measured in arbitrary unit (i.e. spatial length in unit of lattice constant) including the temperature *T* which is in reduced units of the Boltzmann constant.

Simulations are performed on a $550^3$ lattice for a sufficiently long time ($10^7$) steps with a number of independent samples (*100–1000*) over a wide range of temperatures. Different sample sizes are also used to verify the reliability of the qualitative trends from our data presented here. A number of local and global physical quantities such as radius of gyration, root mean square displacement of the center of mass, structure factor, contact map, etc. are examined as a function of temperature. The conformation of the protein exhibits a monotonous response from a random-coil of folded (globular) segments in native phase to tenuous fibrous conformations on raising the temperature; it exhibits a non-monotonic response with a re-entrant conformation involving enhanced globularity before reaching a steady-state conformation on further heating. While most segmental folds disappear in denatured phase while some persist even at a very high temperature (see below).

Figure 1 shows the variation of the average radius of gyration ($R_g$) with the temperature. At low temperatures (*T=0.010 – 0.015*), the radius of gyration remains almost constant with its lowest magnitude ($R_g \sim 22.5$) in its native phase. Unlike many proteins (globular in native phase), COVN appears to be expanded into a random coil (see below) signature of an intrinsically disordered [7] protein. Raising the temperature (*T = 0.015 – 0.023*) leads to a monotonic increase to its maximum *Rg ~54.64±2.60* at *T= 0.0230*. On further heating, the radius of gyration decreases sharply in a narrow range of temperature (*T = 0.023 – 0.025*) to a minimum value ($R_g \sim 38.17±1.72$) at *T = 0.0246* before it begins to increase with the temperature (*T= 0.0250 – 0.0268*) again until it reaches a second maximum ($R_g \sim 51.00±2.24$) at *T = 0.0268*. Beyond the second peak, the radius of gyration continues to decay slowly towards its saturation with the temperature in denatured phase ($R_g \sim 41.4±2.17$ at *T=0.032*, $R_g \sim 38.23±1.96$ at *T=0.050*). Note that this trend is clear despite a relatively large fluctuation in data. To our knowledge, we are not aware of such a non-linear thermal response of such proteins.

Representative snapshots (figure 1, see also figure S1) of the protein at selected temperatures shows the variations in nature of the self-organizing structures over the range of temperature. For example, in native phase (*T=0.010, 0.014*) we see local segmental folding with a chain of folded blobs in a random-coil-like conformation (see below) in contrast to a global folding one generally expects. Local folds begin to disappear at high temperatures but still persist in smaller sizes. Segmental folds appear to be distributed along the entire protein backbone at both maxima and at high temperatures in denature phase while the segmental folds at the minimum and in native phase are localized.

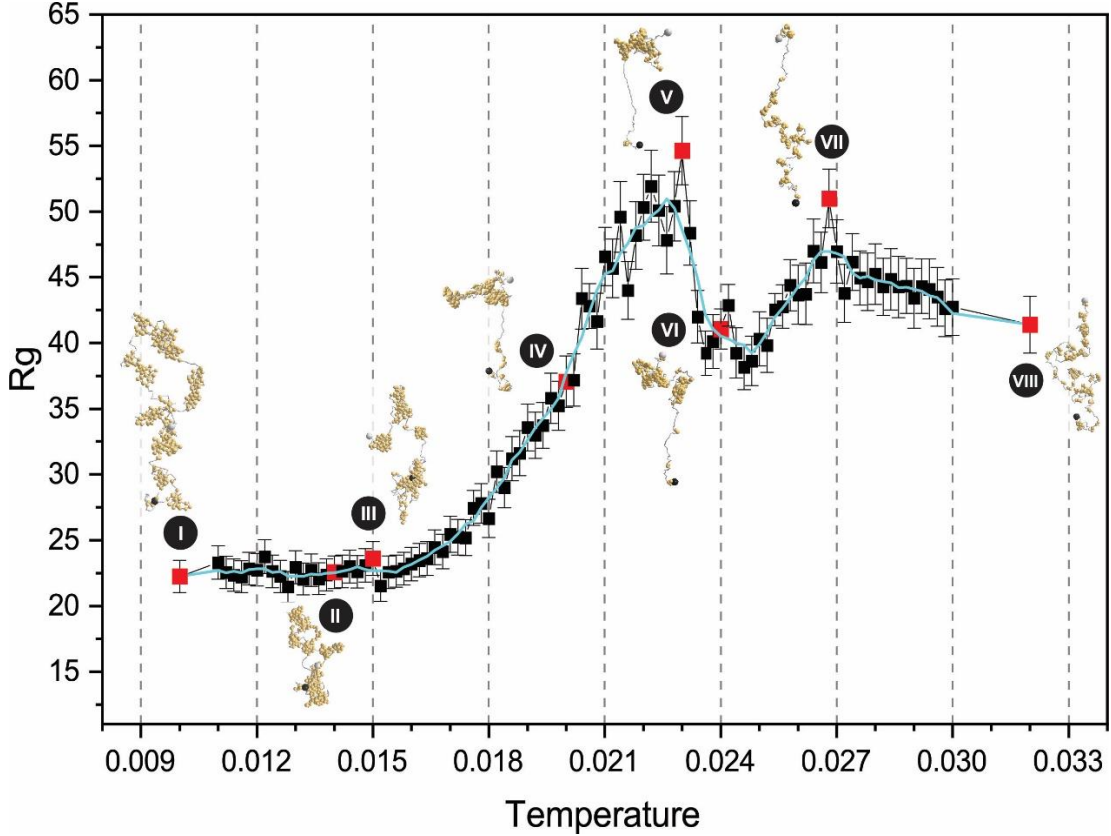

Figure 1: Variation of the average radius of gyration ($R_g$) with the temperature. Some snapshots (at the time step t = $10^7$) are included at representative temperatures: (*i*) *T=0.0100*, (*ii*) *T=0.0140*, (iii) *T=0.0150*, (iv) *T = 0.0200*, (v) *T = 0.0230* (first maximum), (vi) *T= 0.0240* (minimum), (vii) *T = 0.0268* (second maximum), (viii) *T= 0.0320*. Size of the self-organized segmental assembly represents the degree of globularization. In snapshots, gold spheres represent residues in contact, the large black sphere is the first residue [1]M and large grey sphere is the last [422]A (see figure S1).

How to quantify the distribution of residues over length scales? To assess the mass (distribution), we have analyzed the structure factor $S(q)$ defined as,

$$S(q) = \langle \frac{1}{N} \left| \sum_{j=1}^{N} e^{-i\vec{q}\cdot r_j} \right|^2 \rangle_{|\vec{q}|} \qquad (2)$$

where $r_j$ is the position of each residue and $|q| = 2\pi/\lambda$ is the wave vector of wavelength $\lambda$. Using a power-law scaling $S(q) \propto q^{-1/\gamma}$, one may be able to evaluate the power-law exponent $\gamma$ and estimate the spread of residues over the length scale $\lambda$. Overall size of the protein chain is described by its radius of gyration ($R_g$). Therefore, the structure factor over the length scale comparable to protein size ($\lambda \sim R_g$) can provide an estimate of the effective dimension $D$ of the protein conformation via scaling the number of residues ($N$) $N \propto \lambda^D$ where $D = 1/\gamma$. Variations of $S(q)$ with the wavelength $\lambda$ comparable to radius of gyration of the protein over the entire range of representative temperatures are presented in figure 2.

In the native phase (*T = 0.0150*) where the radius of gyration is minimum (*Rg ~ 22.5*), the effective dimension *D ~ 2.053* of the protein shows that the overall spread is not globular. It is rather random-coil, a chain of segmental globules (see figure 1). In unfolding-transition regime (*T=0.020*), the effective dimension $D \sim 1.726$ decreases while retaining its partial folding towards C-terminal (see below). Continuous increasing the temperature leads to maximum unfolding (*T = 0.0230*) where the protein chain stretches to its maximum gyration radius ($Rg \sim 55$) with lowest effective dimension $D \sim 1.579$ with a couple of unfolded segments (see below). Further heating leads to contraction with a lower radius of gyration ($Rg \sim 22.5, T = 0.0246$) with a higher effective dimension $D \sim 2.389$, which indicates more compact conformation than that in its native phase, a thermal-induced folding. The effective dimension begin to reduce with increasing the temperature further as the protein conformation approaches a tenuous structure, i.e. $D \sim 1.579$ at *T=0.036*.

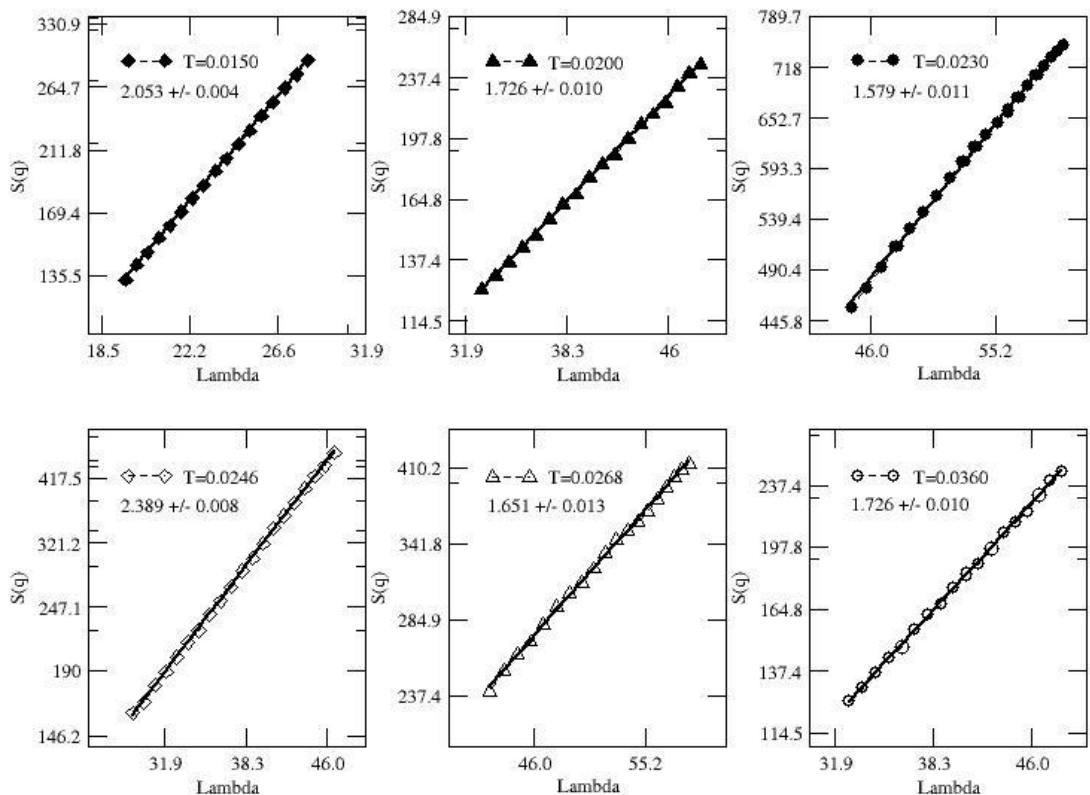

Figure 2: Structure factor *S(q)* versus wavelength (lambda ($\lambda$)) comparable to radius of gyration of COVN on a log-log scale at representative temperatures.

Let us look closer into the local structures by examining the contact map in depth as presented in figure 3 (see also figure S2). First, we notice that the number of residues ($N_r$) within the range of interaction of each residues along the backbone, is higher at lower temperatures. However, the distribution of $N_r$ is highly heterogenous and concentrated towards specific segments ($^{65}$L, $^{110}$Y, $^{224}$L, $^{257}$K, $^{370}$K, $^{374}$K). The degree of folds appears to be significant at these globularization centers (in particular segment $^{367}$T-$^{380}$A) even at higher temperatures although it is highest in native to denature transition region (see also figure 1). Thermal response of contact profiles of each center of folding appears similar except $^{65}$L which exhibits a non-linear (somewhat oscillatory) response (see the right section of figure 3).

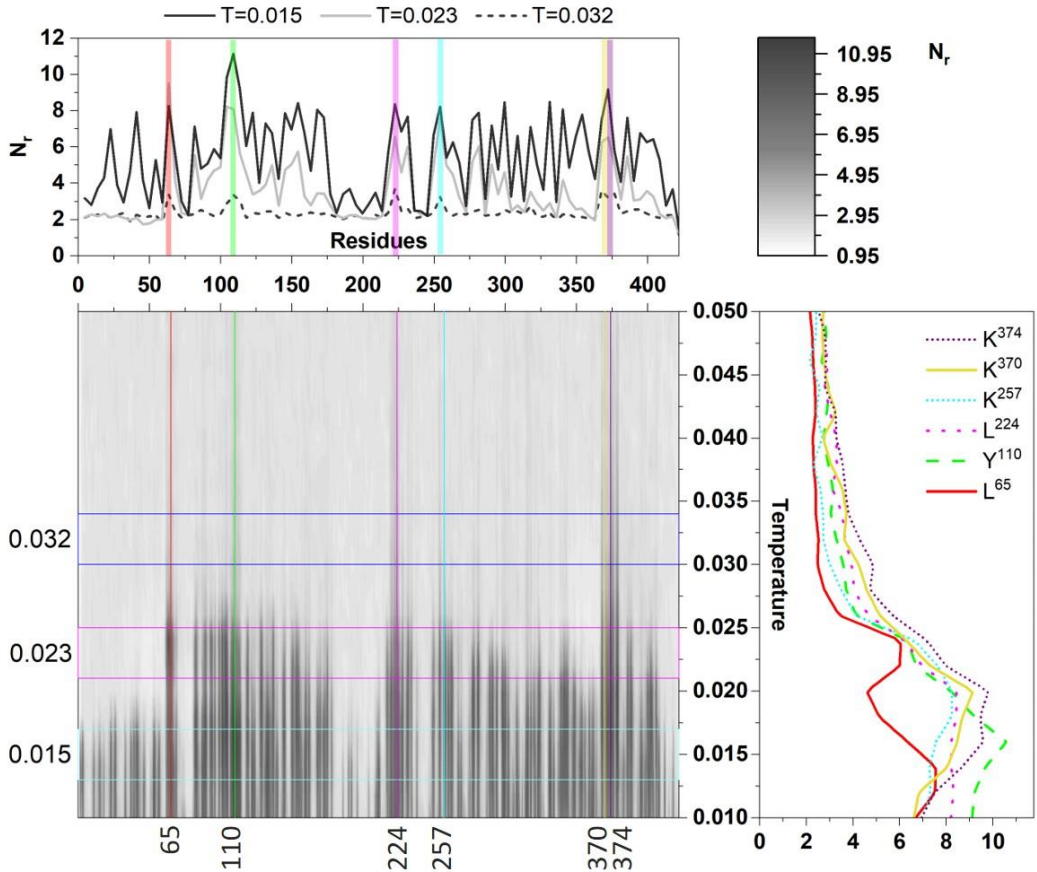

Figure 3: Average number ($N_r$) of residues in contact along the backbone of COVN as a function of temperature. Top figure shows the contacts at representative temperatures in a native phase ($T=0.015$), at the first (maximum) peak of the radius of gyration ($T = 0.0230$), and in a highly denatured phase ($T=0.0320$). These regions of marked in the center three dimensional figure with the scale at the upper right corner. Right figure shows the thermal response of the contact profile of specific centers of folding.

Thus, the thermal response of COVN protein is non-linear with a random coil of folded blobs in native phase to a systematic unfolding, refolding, and unfolding as the protein denatures on increasing the temperature. The radius of gyration increases on raising the temperature, first monotonically from a minimum in its native state to a maximum value. Further heating leads to a sharp decline (the protein contracts) in a narrow temperature range followed by increase (protein expands) again to a second maximum with a local minimum in between. The radius of gyration at the local minimum is larger than that in its native state but the segmental globularization is localized towards the second half (C-terminal) while the first half (N-terminal) of the protein acquire a fibrous configuration. Continued heating causes COVN to approach a steady-state value with a small contraction rate.

Scaling analysis of the structure factor is critical in quantifying the overall spread of COVN by evaluating its effective dimension D. In native phase, $D \sim 2.053$ ($T=0.0150$, native phase), $D \sim 1.716$ ($T=0.0200$, intermediate denature phase), $D \sim 1.579$ ($T=0.0230$, first maximum), $D \sim 2.389$ ($T=0.0246$, local minimum), $D \sim 1.651$ ($T=0.0268$, second maximum), $D \sim 1.726$ ($T=0.0360$, denatured). These estimated are consistent with the thermal response of the radius of gyration. Active zones of folded segments are identified from a detailed analysis of the contact map profile where the degree of folding can be quantified from the average contact measures. Segmental denaturing around residues such as $^{65}$W, $^{110}$Y, $^{224}$L, and $^{374}$P by technique other than thermal agitations may eradicate the specific functionality of COVN.

**Acknowledgement:** Support from the Chulalongkorn University Dusadi Phipat scholarship award to Warin Rangubpit has been instrumental for her visit to University of Southern Mississippi. We thank Brian Olson for helping with computer support. The authors acknowledge HPC at the University of Southern Mississippi supported by the National Science Foundation under the Major Research Instrumentation (MRI) program via Grant # ACI 1626217.

## Supplementary material

## Thermal-induced unfolding-refolding of a nucleocapsid COVN protein

Warin Rangubpit[1,2], Pornthep Sompornpisut[2], R.B. Pandey[1*]

[1]*School of Mathematics and Natural Sciences, University of Southern Mississippi, Hattiesburg, MS 39406, USA*
[2]*Center of Excellence in Computational Chemistry, Department of Chemistry, Chulalongkorn University, Bangkok 10330, Thailand*

Few representative snapshots shows the effect of temperature on folding.

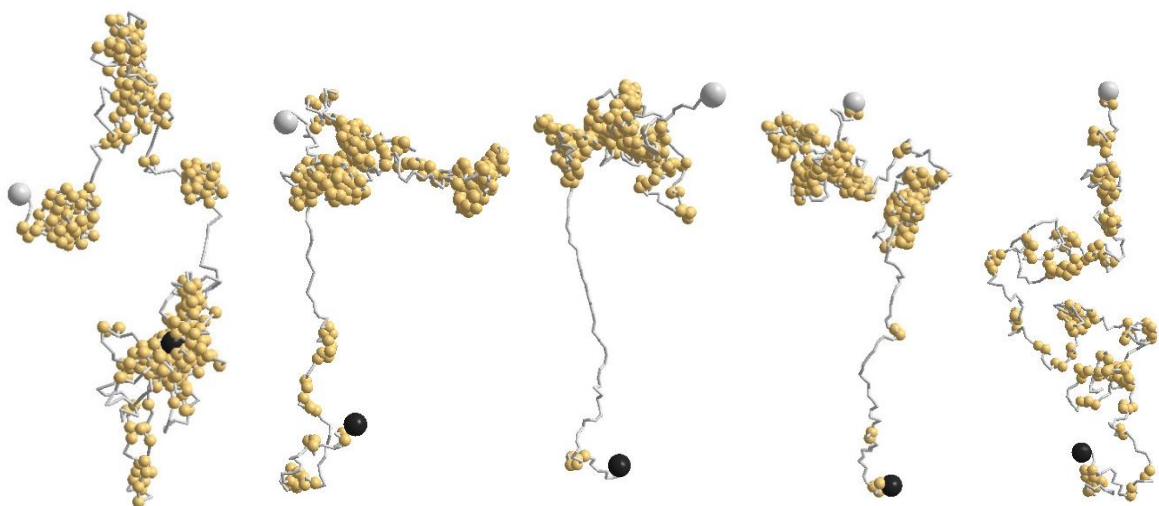

Figure S1: Snapshots of the protein conformation at time step $10^7$ at representative temperatures *T=0.0150, T = 0.0200, T = 0.0230* (first maximum $R_g$), *T= 0.0240* (minimum $R_g$), (vii) *T = 0.0268* (second maximum $R_g$) from left to right. Gold spheres represent residues in contact, the large black sphere is the first residue $^1$M and large grey sphere is the last $^{422}$A.

Details of contact map presented in figures S2 and S3 (for entire protein) shows how the degree of folding in different segments in native phase (*T=0.0150*) decreases with unfolding on raising the temperature. For example, the degree of folding in segments $^{118}$P–$^{164}$Q and $^{276}$G–$^{344}$G reduces dramatically by increasing the temperature *T = 0.0150 – 0.0230* as the protein chain denatures to its maxiumum extension (*T=0.0230*). Further heating (*T=0.0230 – 0.0246*) leads to eradicating a large fraction of these folds while the remaining folds in specific segments e.g. $^{62}$K –$^{70}$G, $^{81}$P–$^{114}$L, $^{226}$T–$^{239}$A, $^{248}$T–$^{270}$N, $^{367}$T–$^{380}$A appear to induce contraction in spread of the protein. The degree of folding reduced on continued heating but the persistence of some folded segments e.g. $^{367}$T–$^{380}$A (along with other segments with comparably low folds) lead to futher expansion of the protein (*T=0.0268*) before it reaches to a stable conformation.

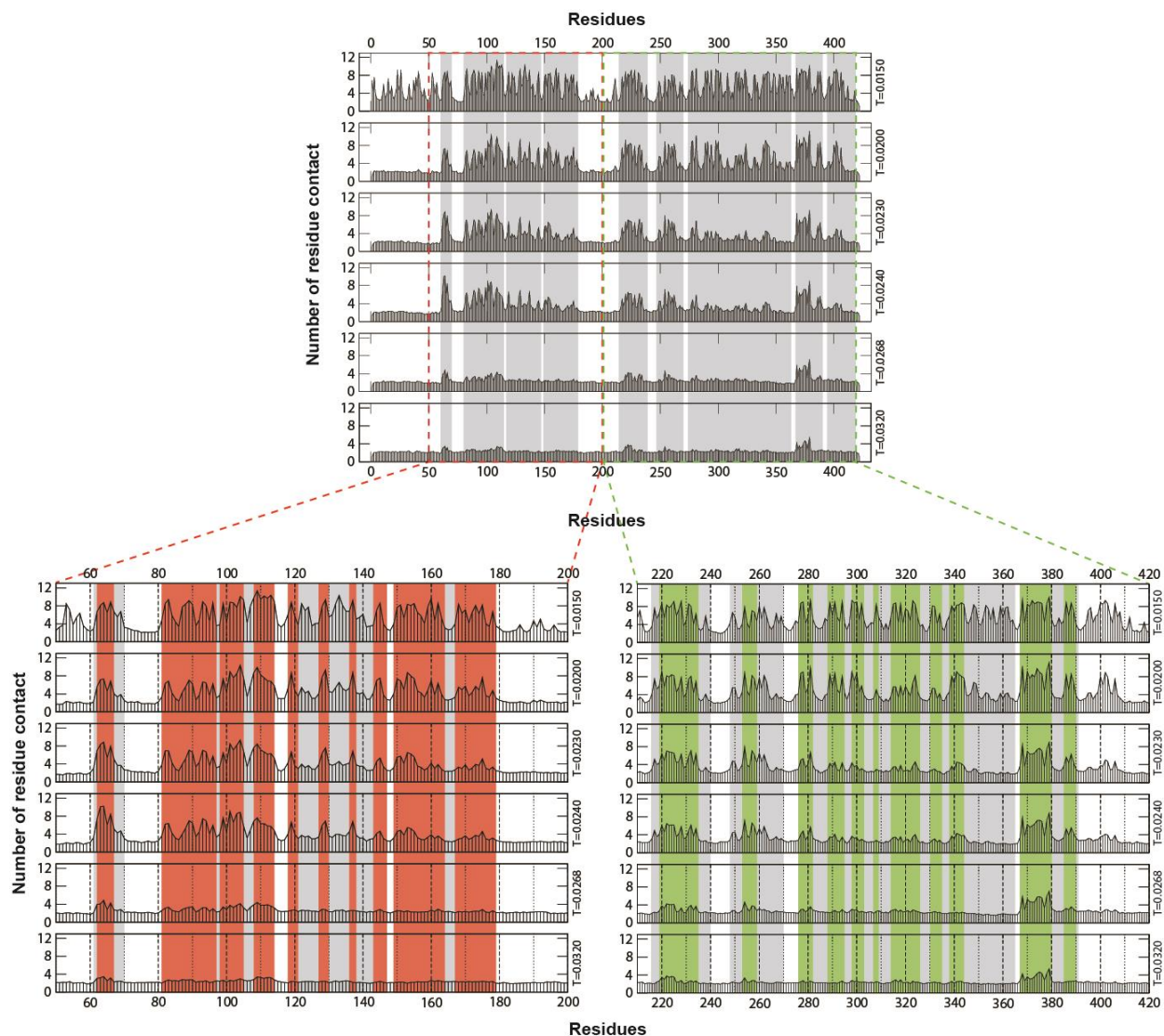

Figure S2: Average number $N_n$ contacts of each residue at representative temperatures $T=0.0150 - 0.032$.

Following segments of COVN with relatively high degree of folding are:
Sequences in left figure: $^{62}$K–$^{67}$F, $^{81}$P–$^{97}$G, $^{98}$G–$^{105}$L, $^{108}$R–$^{114}$L, $^{118}$P–$^{121}$S, $^{127}$N–$^{130}$G, $^{136}$T–$^{138}$G, $^{143}$P–$^{147}$I, $^{149}$T–$^{164}$Q, $^{167}$T–$^{179}$G. Sequences in right figure: $^{219}$E–$^{235}$A, $^{253}$A–$^{259}$P, $^{276}$G–$^{282}$Q, $^{288}$G–$^{295}$Q, $^{298}$D–$^{303}$P, $^{307}$Q–$^{309}$A, $^{314}$A–$^{326}$T, $^{330}$T–$^{335}$H, $^{338}$I–$^{344}$D, $^{367}$T–$^{380}$A, $^{385}$Q–$^{390}$Q, $^{399}$A–$^{408}$L
Note that the folding remains around some segments, e.g. $^{367}$T–$^{380}$A even at high temperature.